\documentclass{article}
\usepackage{graphicx}
\usepackage{amsmath}
\usepackage{verbatim}
\usepackage{url}
\usepackage{geometry}
\usepackage{lscape}
\usepackage[authoryear, round, sort]{natbib}
\bibpunct{(}{)}{;}{a}{}{,}

\begin{document}
\title{A Novel Sequence-Based Antigenic Distance Measure for H1N1, with Application to Vaccine Effectiveness and the Selection of Vaccine Strains}
\author{$\mbox{Keyao Pan}^1$, $\mbox{Krystina C. Subieta}^3$, and
$\mbox{Michael W. Deem}^{1,2}$\\\\
Department of $^1$Bioengineering and $^2$Physics \& Astronomy, Rice University,\\
6100 Main Street, Houston, TX 77005\\
$^3$Department of Mechanical and Aerospace Engineering, University of Florida,\\
Gainesville, FL 32611}

\date{}

\maketitle

\begin{abstract}
H1N1 influenza causes substantial seasonal illness and was the
subtype of the 2009 influenza pandemic. Precise measures of
antigenic distance between the vaccine and circulating virus strains help researchers design influenza vaccines with high vaccine effectiveness.   We here
introduce a sequence-based method to predict vaccine effectiveness in
humans. Historical epidemiological data show that this
sequence-based method is as predictive of vaccine effectiveness as hemagglutination
inhibition (HI) assay data from ferret animal model studies.
Interestingly, the expected vaccine effectiveness is greater against H1N1
than H3N2, suggesting a stronger immune response
against H1N1 than H3N2. The evolution rate of hemagglutinin in H1N1
is also shown to be greater than that in H3N2, presumably due to greater immune
selection pressure.
\end{abstract}


\section{Introduction}
\label{sec:Introduction}

The annual trivalent vaccine for influenza contains one H3N2 strain,
one H1N1 strain, and one influenza B strain. This vaccine is
currently the primary tool to prevent influenza infection and to
control influenza epidemics. Due to the fast
evolution of the influenza virus, the components of the influenza
vaccine are changed for many flu seasons. Even though the vaccine is
usually redesigned to match closely the newly evolved influenza virus
strains, there occasionally has been a suboptimal match between vaccine
and virus. Partly for this reason, vaccine effectiveness has varied in different
years. The desire to have a vaccine with
high effectiveness makes the prediction of the circulating influenza
strain for the next influenza season a key step in vaccine design.
A goal of the WHO is to recommend vaccine strains for the next flu
season that will have the smallest antigenic distances to the dominant
circulating strains in the next flu season, which often means using
the dominant circulating strains in the current flu season as a reference.

A variety of distance measures have been developed to evaluate the
degree of match between the vaccine strain and the dominant circulating
strain. The hemagglutinin protein (HA) of influenza is primarily focused
upon for this distance calculation since hemagglutinin is the dominant antigen for
protective human antibodies and exhibits the highest evolutionary
rate among all the influenza genes \citep{Rambaut2008}. A widely used definition of antigenic
distance is calculated from hemagglutination inhibition data from
ferret animal model studies. To compare a pair of
strains, a 2-by-2 HI titer matrix is built, and the antigenic
distance is extracted from this matrix. This distance can be further
refined by a dimensional projection technique termed antigenic
cartography \citep{Smith2004}. The mathematical basis of antigenic cartography is the dimension reduction of the shape space in which each point represents an influenza virus strain and the distance between a pair of points represents the antigenic distance between the corresponding strains. Note that antigenic cartography does not yield the distance data itself, but assesses the distance between the given vaccine strain and dominant circulating strain by globally considering the effect of all the strains and the antigenic distances among them. In the original literature of antigenic cartography \citep{Smith2004}, hemagglutination inhibition data were the input of the antigenic cartography algorithm that obtains the final results of distances. Antigenic distances can also be
defined by the amino acid sequences of the strains using
computer-aided methods, in which the fraction of substituted amino acid in the dominant hemagglutinin epitope bound by antibody is defined by $p_\mathrm{epitope}$ as a sequence-based antigenic distance measure \citep{Gupta2006,Pan2009,Deem2009}. The amino acid sequences are downloaded from databases and processed to obtain
these distance measures. The $p_\mathrm{epitope}$ sequence-based
method has been shown to be an effective antigenic distance measure
between two strains of H3N2 \citep{Deem2003,Gupta2006,Pan2009}.  To be clear, antigenic distance is a quantity that should define difference of viral strains, as determined by the human immune system.  Ferret HI data are not the only or even the best measure of antigenic distances.

The vaccine effectiveness, which varies from year to year, correlates
with the antigenic distance between the vaccine strain and the dominant
circulating strain. Thus the vaccine effectiveness can be predicted by
calculating the antigenic distance. Such \emph{a priori} estimation
of the vaccine effectiveness guides health authorities to determine the
appropriate strain for the vaccine component for the coming flu
season. For H3N2 influenza, the $p_\mathrm{epitope}$ method offers a
prediction of vaccine effectiveness that has a higher correlation
coefficient with vaccine effectiveness in humans than do distances
derived by other methods \citep{Gupta2006,Pan2009}. In this paper,
we develop the $p_\mathrm{epitope}$ method for H1N1 influenza. In
Materials and Methods we describe the
epidemiological data used to calculate vaccine effectiveness and the
animal model or sequence data used to calculate antigenic distance.
In Results we show the correlation of antigenic
distance with vaccine effectiveness. We discuss the results in the Discussion.

\section{Materials and Methods}
\label{sec:Materials_and_Methods}

\subsection{Identities of Vaccine Strains and Dominant Circulating Strains}

The vaccine strain selection by WHO in each year follows a standard procedure. The vaccine strains are reviewed every year and are usually changed every two to three years. We used the H1N1 vaccine strains and H1N1 dominant circulating strains in the epidemiological literature that provided vaccine effectiveness data used in this study.

\subsection{Estimation of Vaccine Effectiveness}

The H1N1 vaccine
effectiveness is gathered from epidemiological literature regarding
the influenza-like illness rate of unvaccinated ($u$) and vaccinated
people ($v$). Vaccine effectiveness can be described by the following definition:
\begin{equation}\label{eq:efficacy}
\mbox{vaccine effectiveness} = \frac{u-v}{u}.
\end{equation}

To calculate vaccine effectiveness and its standard error, we let
$N_u$ and $N_v$ denote the number of subjects in the unvaccinated
and vaccinated group, $n_u$ and $n_v$ denote the number of illness in the unvaccinated and vaccinated group, respectively. The values and the standard errors of $u$, $v$,
and vaccine effectiveness are
\begin{eqnarray}
\label{eq:valu} u &=& n_u / N_u\\
\label{eq:valv} v &=& n_v / N_v\\
\label{eq:valve} \mathrm{VE} &=& \frac{u-v}{u} = \frac{n_u N_v - n_v N_u}{n_u N_v}\\
\label{eq:erroru} \sigma_u &=& \sqrt{\frac{u\left(1-u\right)}{N_u}}\\
\label{eq:errorv} \sigma_v &=& \sqrt{\frac{v\left(1-v\right)}{N_v}}\\
\label{eq:errorve} \sigma_\mathrm{VE} &=& \left(\frac{v}{u}\right)
\sqrt{\left(\frac{\sigma_v}{v}\right)^2 +
\left(\frac{\sigma_u}{u}\right)^2} =
\sqrt{\left(\frac{1}{u}\right)^2 \sigma_v^2 +
\left(\frac{v}{u^2}\right)^2 \sigma_u^2}.
\end{eqnarray}
If the vaccine effectiveness is averaged from $N$ studies,
$\sigma_\mathrm{VE}^2 = \left(\sum_i \sigma_{\mathrm{VE}i}^2\right)
/ N^2$ where $\sigma_{\mathrm{VE}i}$ is the standard error of the
$i$--th study.

Compared to H3N2, subtype H1N1 viruses were dominant in fewer years. Based on the proportions of samples of H3N2, H1N1, and influenza B collected in each year during 1977--2009, widespread H1N1 circulation was observed in approximately 10 seasons. Epidemiological studies on vaccine effectiveness were absent for some years when H1N1 circulated. Additionally, we used the criteria listed below to filter all available literature.

To ensure that the vaccine effectiveness we collected from the literature is for
H1N1, the seasons and the geographic regions of the epidemiological
studies in the literature were compared with the influenza activity
information in WHO Weekly Epidemiological Records
to confirm that those regions were dominated by H1N1 in those
seasons. Subjects were restricted to 18--64 year old healthy adult
humans to avoid effects of an underdeveloped immune system in
children or of immunosenescence in senior people. If more than one measure of
vaccine effectiveness was collected for the same season, they were
averaged to minimize the statistical noise.


In order to minimize the effect on vaccine effectiveness from co-circulating subtypes such as H3N2, only the epidemiological data collected in the regions and in the flu seasons in which the H1N1 subtype was dominant were applied to calculate the vaccine effectiveness in this study. The seasons in which the H1N1 subtype was dominant were reported by the literature on H1N1 vaccine effectiveness.  The studies cited in Table \ref{tab:summary} for the calculation of vaccine effectiveness gave the subtype of the predominant epidemic virus as well as of the virus sampled from the subjects with influenza-like illness (ILI). In addition, the dominance of H1N1 subtype is also available in the CDC Morbidity and Mortality Weekly Reports and the WHO Weekly Epidemiological Record. For the data in Table \ref{tab:summary}, the dominance of H1N1 subtype was shown in these references.


The vaccine effectiveness collected from various flu seasons and regions were measured with standard errors. Biases in the vaccine effectiveness are due to the complexity of the vaccine effectiveness measurement, including the character of the human population studied, such as age, immune history, and health condition; the influence of co-circulating H3N2 influenza strains; the character of the vaccine distributed, such as live attenuated virus vaccine, inactivated split-virus vaccine produced by virion disassembly, or subunit vaccine only containing hemagglutinin and neuraminidase; the method of epidemiological measurement of influenza infection, such as virus detection, confirmed symptomatic influenza, or influenza-like illness (ILI); the design of the experiment, such as natural infection or experimental challenge study; and the progression of the epidemic in the population under study. These biases are thus inevitable with current technology. Here, we applied the following methods to minimize biases in the vaccine effectiveness data. Subjects in the studies were confined to 18--64 years old healthy adult humans to preclude the interference of the feeble immune system in children or in senior people, because variation in the capability of the immune system is a determinant of the vaccine effectiveness given the same pair of vaccine strain and dominant circulating strain. Only epidemiological studies in the season and the region in which H1N1 subtype was dominant were used to obtain the vaccine effectiveness data. The vaccine involved in the referred studies is an inactivated vaccine. Other types such as cold-adapted nasal spray vaccine were excluded. The epidemiological measurement of infection in all the referred studies used ILI as the criterion. Not all studies designed the experiment as a challenge study. We assume that the epidemic propagates in the population in a similar way in each season.  These criteria are used to filter the available references and to obtain vaccine effectiveness data with minimum bias. The standard errors of the data are presented here. These criteria reduced the number of practical references for each season. Our metaanalysis considered 50 peer-reviewed papers, all we could find in the literature.  We list the ones that satisfy our selection criteria for each of the years, typically 1--3 \emph{per year}.

\subsection{Antigenic Distance Measured By Sequence Data}

\begin{figure}
\centering
\includegraphics[width=5in]{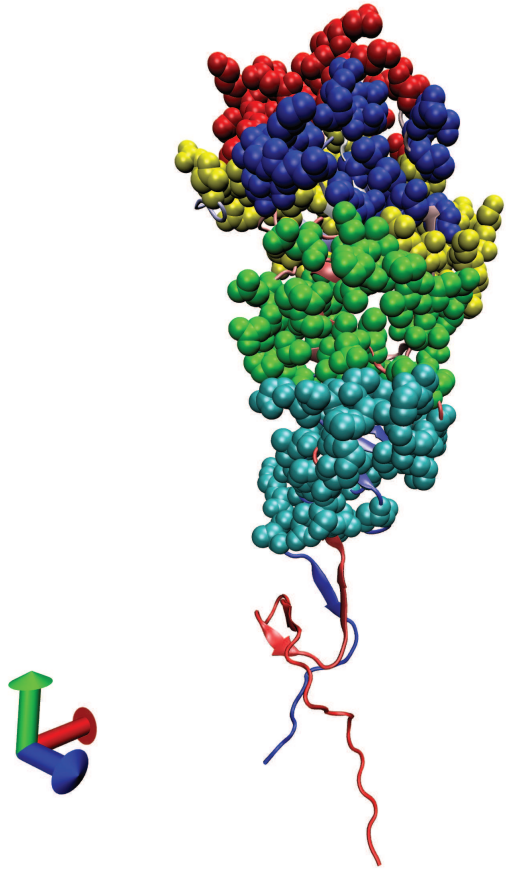}
\caption{HA1 domain of the H1 hemagglutinin in the ribbon format
(PDB code: 1RU7). Epitope A (blue), B (red), C (cyan), D (yellow),
and E (red) are space filling. These five H1 epitopes are the analogs of the well-defined H3 epitopes \citep{Deem2009}.}
\label{fig:epitopes}
\end{figure}

Figure \ref{fig:epitopes} shows the HA1 domain with five epitopes of the H1 subtype hemagglutinin. As the improvement of a previous definition of H1 epitopes \citep{Caton1982}, these five H1 epitopes are recognized by host antibodies and are identified by mapping the well-defined epitopes in H3 hemagglutinin \citep{Wiley1981,Macken2001} to H1 hemagglutinin and using sequence entropy to find additional sites under selection \citep{Deem2009}.

The antigenic distance between the vaccine strain and the dominant circulating strain is the input for the vaccine effectiveness prediction. The fraction of mutated amino acids in the epitope region of HA, or the $p$-value, is an antigenic distance measure to quantify the similarity between two strains \citep{Gupta2006}.
One $p$-value is calculated for each H1 epitope
\begin{equation}\label{eq:pvalue}
p\mbox{-value} = \frac{\mbox{number of mutations in the
epitope}}{\mbox{number of amino acids in the epitope}}.
\end{equation}
The $p_\mathrm{epitope}$ is defined as the maximum of five
$p$-values for the five epitopes, and the dominant epitope is
defined as the corresponding epitope. This definition, i.e. assumption, has lead for H3N2 to vaccine effectiveness predictions that correlate with those observed \citep{Gupta2006}.

Another sequence-based antigenic distance measure uses the fraction of mutated amino acid in all the five epitopes
\begin{equation}\label{eq:pallepitope}
p_{\text{all-epitope}} = \frac{\mbox{number of mutations in all the five
epitopes}}{\mbox{number of amino acids in all the five epitopes}}.
\end{equation}

As an alternative to $p_\mathrm{epitope}$ and $p_{\text{all-epitope}}$, $p_\mathrm{sequence}$ is
also used with the definition
\begin{equation}\label{eq:psequence}
p_\mathrm{sequence} = \frac{\mbox{number of mutations in the HA1
domain of hemagglutinin}}{\mbox{total number of amino acids in the
HA1 domain of hemagglutinin}}.
\end{equation}

\subsection{Antigenic Distance Measured by Hemagglutination Inhibition}

The animal model method to determine the distance between the
vaccine strain and the dominant circulating strain employs the HI assay to
give the HI table. See Table \ref{tab:HItable}: Here $H_{ij}$,
$i,j=1,2$ are four HI titers measuring the capability of antibody
$j$ to inhibit hemagglutinin $i$. Note that in reality, health authorities including WHO and CDC provide HI tables with at least eight antisera to evaluate the antigenic distance between candidate vaccine strains and dominant circulating strain. These HI tables are mathematically equivalent to several $2\times2$ HI tables each of which defines the antigenic distance between one pair of strains in the original HI table. For each pair of strains, we picked up four entries determined by the identities of these two strains and the two corresponding antisera from the original HI table. The $2\times2$ HI tables in this manuscript are used to elaborate the formulae for $d_1$ and $d_2$. In this context Strain 1 is the
vaccine strain and Strain 2 is the dominant circulating strain. Two distance
measures have been derived from these four HI titers in the HI table
\citep{Smith1999,Lee2004}:
\begin{eqnarray}
\label{eq:HIassay1} d_1 &=& \log_2 \left(\frac{H_{11}}{H_{21}}\right)\\
\label{eq:HIassay2} d_2 &=& \sqrt{\frac{H_{11} H_{22}}{H_{21}
H_{12}}}.
\end{eqnarray}
Note that antigenic cartography is carried out on the asymmetrical
distance, $d_1$ \citep{Smith2004}. When the vaccine strain and the dominant
circulating strain in one season were not identical, we searched the
literature for the HI tables with these two strains. The $d_1$ and
$d_2$ values were averaged if multiple HI tables were found for one
season.

\begin{table}
\caption{HI table with two strains and four HI titers.} \centering
\begin{tabular}{c|c c}
\multicolumn{2}{c}{} \\
         & Ferret antisera  & Ferret antisera \\
         & against Strain 1 & against Strain 2 \\\hline
Strain 1 & $H_{11}$         & $H_{12}$ \\
Strain 2 & $H_{21}$         & $H_{22}$ \\
\end{tabular}
\label{tab:HItable}
\end{table}

\begin{landscape}
\begin{table}
\caption{Summary of results. Nine pairs of vaccine strains and dominant
circulating strains in seven flu seasons in the Northern hemisphere
were collected from literature. The quantities $n_u$, $N_u$, $n_v$, $N_v$, $p_\mathrm{epitope}$, $p_{\text{all-epitope}}$, $p_\mathrm{sequence}$, $d_1$, and $d_2$ are defined in Materials and Methods. Only those seasons when H1N1 virus
was dominant in at least one country or region where vaccine
effectiveness data were available were considered. Two different vaccines
have occasionally been adopted in different geographic regions for
the same season, in which case two sets of data were added in this
table. An asterisk signifies that co-circulating H3N2 was also
found in the same country or region in that season; however, the
interference to the final result from H3N2 is expected to be small, and so the sets
of data with a single asterisk were preserved.} \centering
\hbox{
{\tiny
\begin{tabular}{l l l l l l l l l l l l l l}
\\\hline
Season & Vaccine strain & Dominant & Vaccine & $n_u$ & $N_u$ & $n_v$ & $N_v$ & Dominant & $p_\mathrm{epitope}$ & $p_{\text{all-epitope}}$ & $p_\mathrm{sequence}$ & $d_1$ & $d_2$ \\
& & Circulating strain\ddag & effectiveness & & & & & epitope & & & & & \\
& & & (\%) & & & & & & & & & & \\
\hline
1982--83     & A/Brazil/11/78           & A/England/333/80         & $37.0 \pm 12.0 ^ {1}$ & 48 & 118 & 31 & 121$^1$ & A & 0.083 & 0.0311 & 0.0184 & 0$^{10}$    & 1.41$^{10}$ \\
& & & & & & & & & & & & & \\
1983--84     & A/Brazil/11/78           & A/Victoria/7/83          & $38.1 \pm 10.3 ^ {\text{1--3}}$ & 30 & 60 & 21 & 67$^1$ & C & 0.121 & 0.0497 & 0.0337 & 1.13$^{\text{11--13}}$ & 13.66$^{11,13}$ \\
& & & & 55 & 298 & 46 & 300$^2$ & & & & & & \\
& & & & & & & & & & & & & \\
1986--87 (a) & A/Taiwan/1/86            & A/Taiwan/1/86            & $64.8 \pm 14.3 ^ {3,4}$ & 11 & 217 & 13 & 723$^4$ &   & 0 & 0    & 0      & 0    & 1 \\
& & & & & & & & & & & & & \\
1986--87 (b) & A/Chile/1/83             & A/Taiwan/1/86            & $18.5 \pm 12.1 ^ {5}$ & 92 & 878 & 75 & 878$^5$ & B & 0.318 & 0.0807 & 0.0399 & 4$^{\text{12,14--18}}$    & 24.48$^{\text{14,16--18}}$ \\
& & & & & & & & & & & & & \\
1988--89     & A/Taiwan/1/86            & A/Taiwan/1/86            & $43.1 \pm 10.0 ^ {3,5}$ & 119 & 1125 & 89 & 1126$^5$ &   & 0     & 0  & 0    & 0    & 1 \\
& & & & & & & & & & & & & \\
1995--96 (a) & A/Texas/36/91            & A/Texas/36/91            & $60.0 \pm 27.8 ^ {6}$ & 6 & 12 & 2 & 10$^6$ &   & 0  & 0   & 0      & 0    & 1 \\
& & & & & & & & & & & & & \\
1995--96 (b)* & A/Singapore/6/86        & A/Texas/36/91            & $32.2 \pm 5.8 ^ {7}$  & 99 & 652 & 57 & 684$^7$ & A & 0.125 & 0.0559 & 0.0307 & 0.86$^{14,19,20}$ & 2.43$^{14,20}$ \\
& & & & 176 & 652 & 149 & 684$^7$ & & & & & & \\
& & & & & & & & & & & & & \\
2006--07     & A/New Caledonia/20/99    & A/New Caledonia/20/99    & $40.5 \pm 2.5 ^ {8}$  & 1085 & 230729 & 1221 & 436600$^8$ &   & 0     & 0 & 0     & 0    & 1 \\
& & & & & & & & & & & & & \\
2007--08*   & A/Solomon Islands/3/2006 & A/Solomon Islands/3/2006  & $62.8 \pm 12.6 ^ {9}$ & 94 & 262 & 8 & 60$^9$ &   & 0 & 0    & 0      & 0    & 1 \\
\hline
\end{tabular}
}
} \label{tab:summary}
\parbox{7.5in}{\ddag Multiple strains are circulating in each season, while each strain has a specific proportion in the virus population in a certain region and season. The strain with the greatest proportion is defined as the dominant circulating strain, which is listed in this table. The dominant circulating strains in this table were chosen based on the literature on vaccine effectiveness, which also gave the region where the effectiveness data were collected.}
\parbox{7.5in}{Literature used in the metaanalysis: 1. \citep{Couch1986}; 2. \citep{Keitel1988}; 3. \citep{Couch1996}; 4. \citep{Keitel1997}; 5. \citep{Edwards1994}; 6. \citep{Treanor1999}; 7. \citep{Grotto1998}; 8. \citep{Wang2009b}; 9. \citep{Belongia2008}; 10. \citep{Daniels1985}; 11. \citep{Chakraverty1986}; 12. \citep{Smith1999}; 13. \citep{WHOWER1984}; 14. \citep{Hay2001}; 15. \citep{WHOWER1986}; 16. \citep{Kendal1990}; 17. \citep{Donatelli1993}; 18. \citep{Brown1998}; 19. \citep{WHOWER1992}; 20. \citep{Rimmelzwaan2001}.}
\end{table}
\end{landscape}

\section{Results}
\label{sec:Results}



We performed a metaanalysis of identities of the vaccine strains and dominant circulating strains, vaccine effectiveness, and antigenic distances between vaccine strains and dominant circulating strains measured with the HI assay using ferret antisera. In one season dominated by H1N1, epidemiological statistics in a certain region reported in literature was used to fix the values of $n_u$, $N_u$, $n_v$, $N_v$, and the mean and standard error of the vaccine effectiveness.  HI assay data in literature are also used to determine antigenic distance $d_1$ and $d_2$ between the vaccine strain and dominant circulating strain.  Results of the metaanalysis are listed in Table \ref{tab:summary}.  Sequence-based antigenic distances $p_\mathrm{epitope}$, $p_{\text{all-epitope}}$, and $p_\mathrm{sequence}$ are calculated from the sequences of the vaccine strain and dominant circulating strain by equations \ref{eq:pvalue}, \ref{eq:pallepitope}, and \ref{eq:psequence}, respectively. Values of $p_\mathrm{epitope}$, $p_{\text{all-epitope}}$, and $p_\mathrm{sequence}$ in each season dominated by H1N1 are also listed in Table \ref{tab:summary}.

\begin{figure}
\centering
\includegraphics[width=5in]{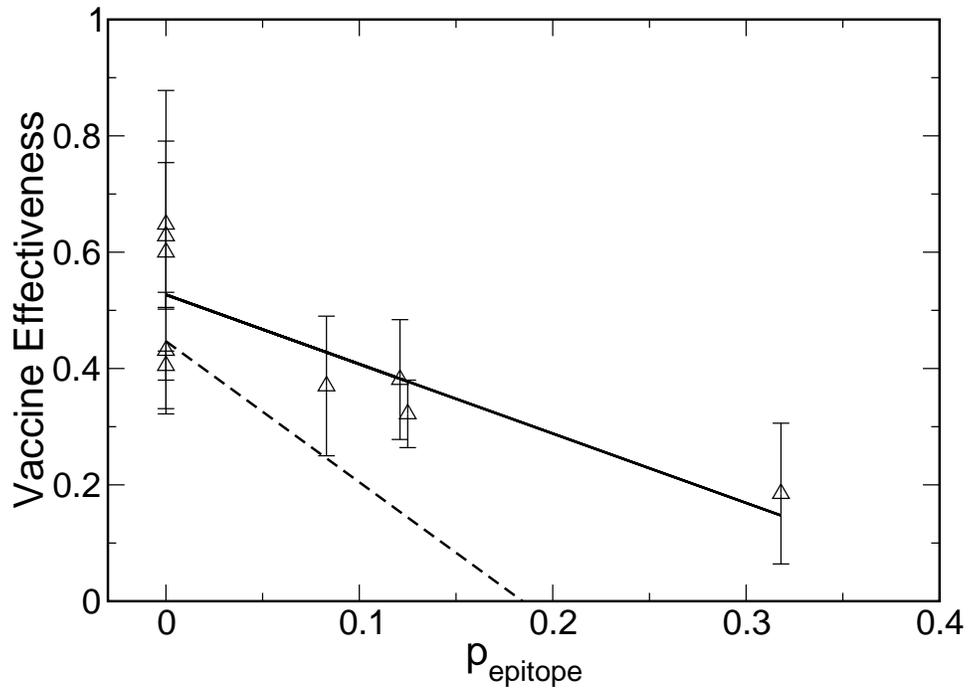}
\caption{Vaccine effectiveness for influenza-like illness correlates with
$p_\mathrm{epitope}$, $R^2=0.68$ (solid line). Data from Table \ref{tab:summary}. The trend line
quantifies vaccine effectiveness as a decreasing linear function of
$p_\mathrm{epitope}$. $\mbox{Vaccine effectiveness} = -1.19 \
p_\mathrm{epitope} + 0.53$. Also shown is the vaccine effectiveness to
H3N2 (dashed line) \citep{Gupta2006}.} \label{fig:pepitope}
\end{figure}

While the number of data points is limited, a linear relationship exists between vaccine effectiveness and $p_\mathrm{epitope}$ by using least squares.  Similar to the case for H3N2 influenza
\citep{Gupta2006}, $p_\mathrm{epitope}$ strongly correlates with
H1N1 vaccine effectiveness, with $R^2 = 0.68$. The fitted model predicts a
vaccine effectiveness of 52.7\% when $p_\mathrm{epitope}=0$, and vaccine
effectiveness is greater than zero when $p_\mathrm{epitope}<0.442$. In
Figure \ref{fig:pepitope}, the fitted trend line is within one standard error of all data points with $p_\mathrm{epitope}>
0$, validating the ability of the $p_\mathrm{epitope}$ model to
predict the vaccine effectiveness with only the sequences of the vaccine
strain and the dominant circulating strain.

Although statistical errors exist in the observed vaccine effectiveness, the collected vaccine effectiveness data reject the null hypothesis that the vaccine effectiveness is independent of $p_\mathrm{epitope}$. The nine pairs of vaccine strains and dominant circulating strains in Table \ref{tab:summary} have five difference antigenic distances between vaccine strain and dominant circulating strain defined by $p_\mathrm{epitope}$. The nine pairs of strains were thus categorized into group 1--5 with $p_\mathrm{epitope}$ equal to 0, 0.083, 0.121, 0.125, and 0.318, respectively, and the average vaccine effectiveness and standard error were calculated for each group. The vaccine effectiveness differences between these five groups were significant, such as group 1 and group 4 ($p = 0.0079$) and group 1 and group 5 ($p = 0.0054$). Moreover, statistical analysis shows that the introduction of $p_\mathrm{epitope}$ is valuable in the selection process of vaccine strains. The slope of the fit line is significantly smaller than zero ($p = 0.0027$). Hence the linear model is able to predict the vaccine effectiveness with the knowledge of $p_\mathrm{epitope}$. In other words the non-zero slope of vaccine effectiveness as a function of $p_\mathrm{epitope}$ is significant to the level of $0.27\%$.

\begin{figure}
\centering
\includegraphics[width=5in]{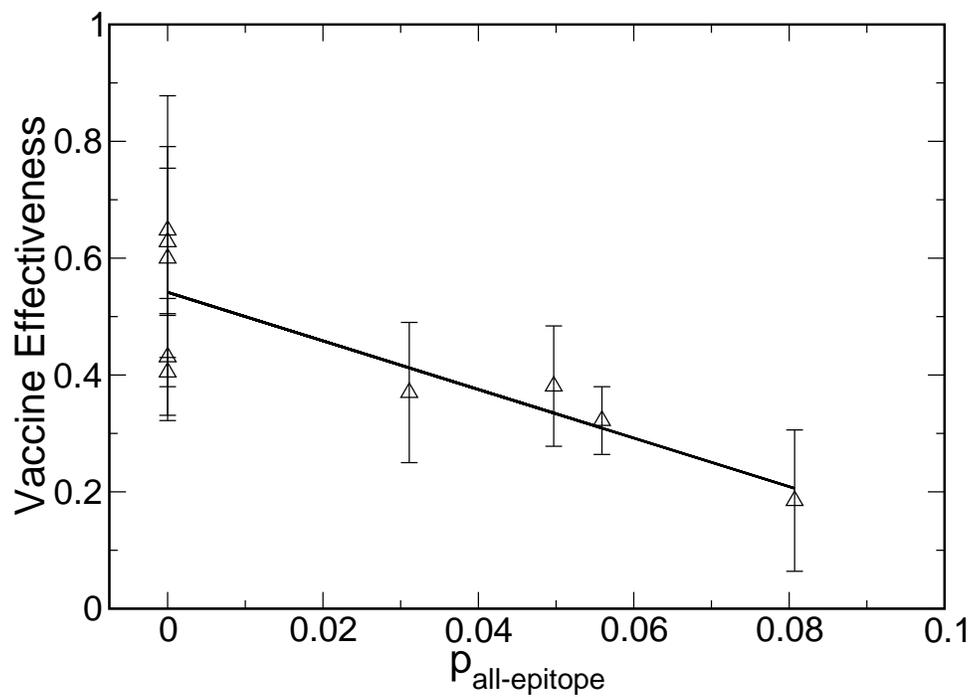}
\caption{Vaccine effectiveness for influenza-like illness correlates with
$p_{\text{all-epitope}}$ with $R^2=0.70$. Data from Table \ref{tab:summary}. The trend line quantifies
vaccine effectiveness as a decreasing linear function of
$p_{\text{all-epitope}}$. $\mbox{Vaccine effectiveness} = -4.16 \
p_{\text{all-epitope}} + 0.54$.} \label{fig:pallepitope}
\end{figure}

\begin{figure}
\centering
\includegraphics[width=5in]{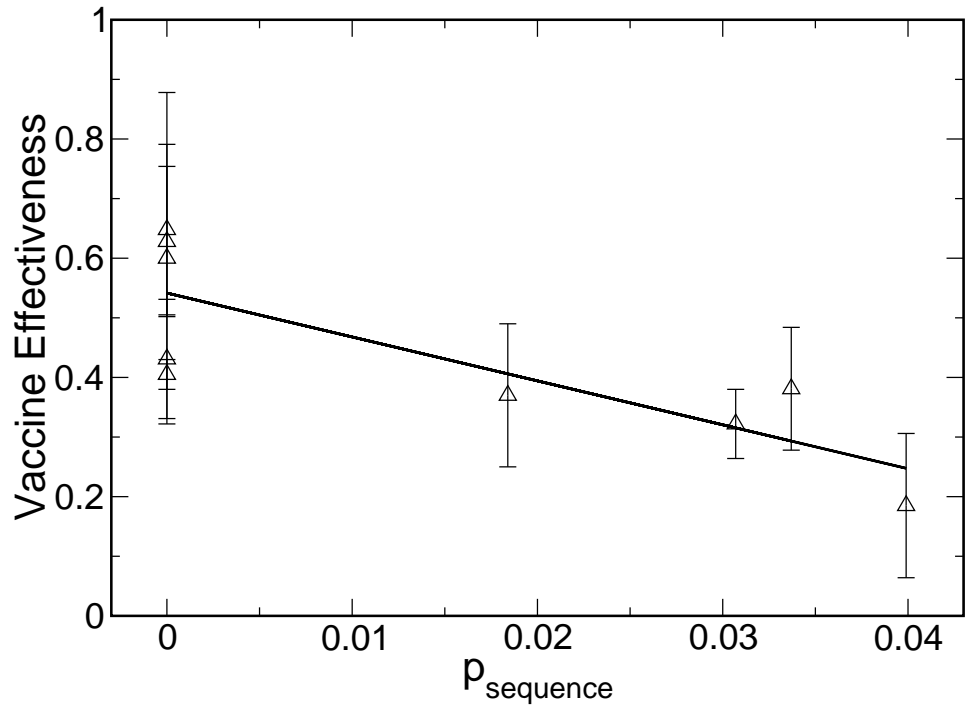}
\caption{Vaccine effectiveness for influenza-like illness correlates with
$p_\mathrm{sequence}$ with $R^2=0.66$. Data from Table \ref{tab:summary}. The trend line quantifies
vaccine effectiveness as a decreasing linear function of
$p_\mathrm{sequence}$. $\mbox{Vaccine effectiveness} = -7.37 \
p_\mathrm{sequence} + 0.54$.} \label{fig:psequence}
\end{figure}

Two other sequence-based antigenic distance measures alternative to $p_\mathrm{epitope}$ are $p_{\text{all-epitope}}$ and
$p_\mathrm{sequence}$. Unlike $p_\mathrm{epitope}$, which focuses
upon the mutations in the antibody binding regions, $p_{\text{all-epitope}}$
calculates the fraction of mutated amino acids
in all the five epitopes, and $p_\mathrm{sequence}$ calculates the fraction of mutated amino acids in the whole HA1 domain of hemagglutinin. The $p_\mathrm{sequence}$
measure is also one of the optional distance measures for
phylogenetic softwares. In Figure \ref{fig:pallepitope}, the correlation
between H1N1 vaccine effectiveness and $p_{\text{all-epitope}}$ has
$R^2=0.70$.
In Figure \ref{fig:psequence}, the correlation
between H1N1 vaccine effectiveness and $p_\mathrm{sequence}$ has
$R^2=0.66$. The predicted 54\% vaccine effectiveness when
$p_{\text{all-epitope}}$ in Figure \ref{fig:pallepitope} and when
$p_\mathrm{sequence}=0$ in Figure \ref{fig:psequence} are almost the same as the 53\% predicted by the $p_\mathrm{epitope}$
method. By contrast $p_{\text{all-epitope}}$ and $p_\mathrm{sequence}$ for H3N2 have less impressive
correlations with H3N2 vaccine effectiveness \citep{Sun2006,Gupta2006}, and $p_{\text{all-epitope}}$ and $p_\mathrm{sequence}$ are
not as effective as $p_\mathrm{epitope}$ as antigenic distance
measures and vaccine effectiveness predictors for H3N2.

\begin{figure}
\centering
\includegraphics[width=5in]{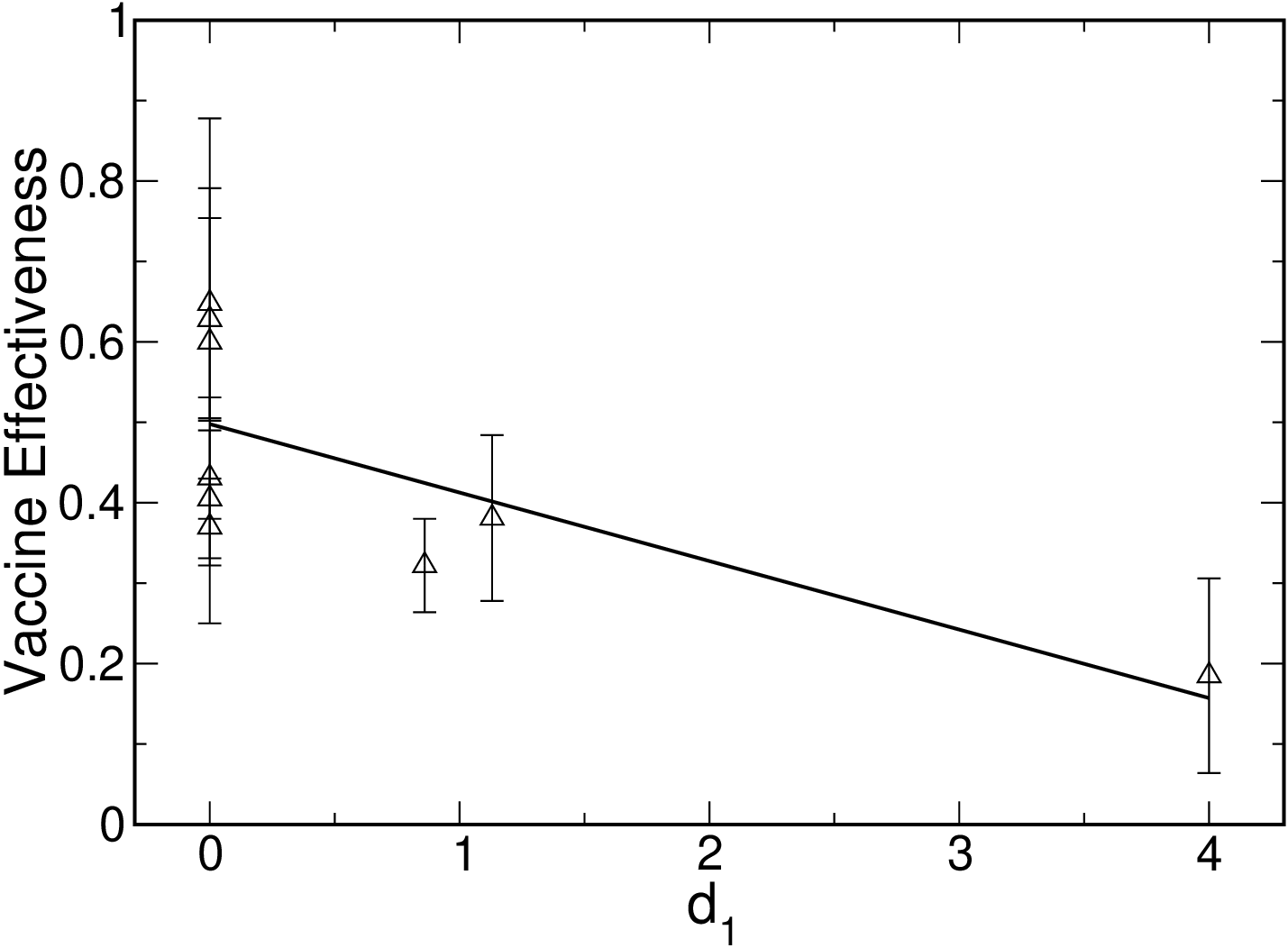}
\caption{The correlation with $R^2=0.53$ between vaccine effectiveness
for influenza-like illness and $d_1$, the antigenic distance defined
by HI assay using ferret antisera. Data from Table \ref{tab:summary}. The $d_1$ values were averaged if
multiple HI assay experimental data were found. The trend line
quantifies vaccine effectiveness as a decreasing linear function of
$d_1$. $\mbox{Vaccine effectiveness} = -0.085 \ d_1 + 0.50$.}
\label{fig:d1}
\end{figure}

\begin{figure}
\centering
\includegraphics[width=5in]{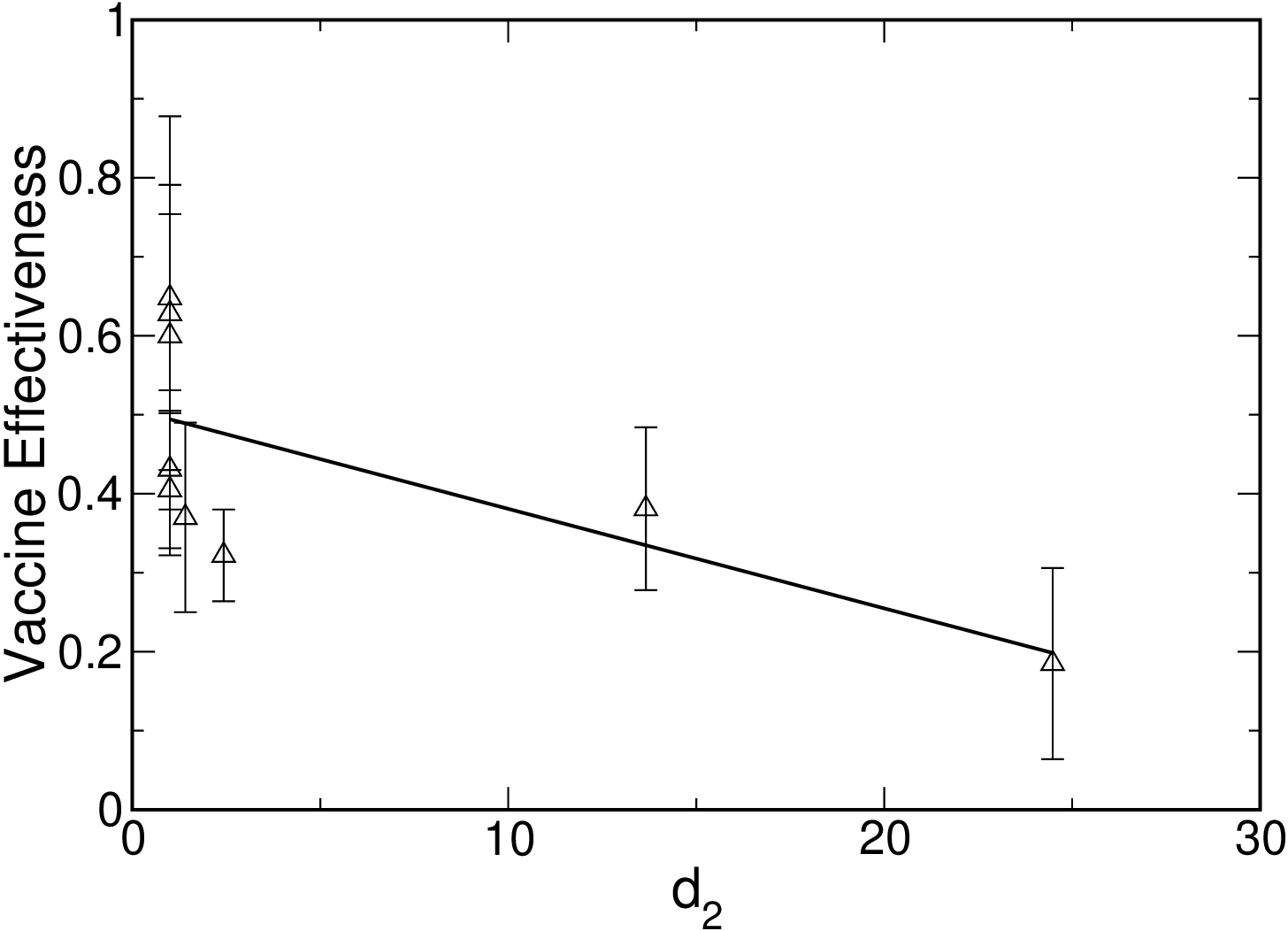}
\caption{The correlation with $R^2=0.46$ between vaccine effectiveness
for influenza-like illness and $d_2$, the antigenic distance defined
by HI assay using ferret antisera. Data from Table \ref{tab:summary}. The $d_2$ values were averaged if
multiple HI assay experimental data were found. The trend line
quantifies vaccine effectiveness as a decreasing linear function of
$d_2$. $\mbox{Vaccine effectiveness} = -0.013 \ d_2 + 0.51$.}
\label{fig:d2}
\end{figure}

The HI assay and derived distance measures $d_1$ and $d_2$ are still
the most widely used measures by researchers and health authorities
to identify newly collected circulating strains. These methods are
used to recommend the vaccine strain for the coming flu season
\citep{WHO2003,WHO2007,WHO2008}, to draw the antigenic map
\citep{Smith2004}, and to support the phylogenetic data
\citep{WHO2003}.
Figure \ref{fig:d1} and \ref{fig:d2} describe the correlation between
vaccine effectiveness and antigenic distances $d_1$ and $d_2$ from the HI assay. A correlation is found in both
figures. In the season 1995--96 in Israel, the vaccine strain is
A/Singapore/6/86 (H1N1) and the dominant circulating strain is A/Texas/36/91
(H1N1), between which the averaged $d_1$ is 0.86. Since the vaccine
effectiveness is only 32.2\%, its discrepancy to the corresponding
effectiveness 42.5\% in the trend line is much larger than one standard
error of vaccine effectiveness. Similarly, the same pair of vaccine
strain and dominant circulating strain introduces a data point further from
the trend line if $d_2$ is used as the distance measure. We also notice that two strains could be antigenically identical as measured with HI assay but antigenically distinct as measured with $p_\mathrm{epitope}$. As shown in Table \ref{tab:summary}, in the season 1982--1983, the H1N1 vaccine strain A/Brazil/11/78 and dominant circulating strain A/England/333/80 presented the antigenic distance measured with HI assay $d_1 = 0$ and the sequence-based antigenic distance measure $p_\mathrm{epitope} = 0.083$. The H3N2 vaccine strain and dominant circulating strain showed identical $d_1$ and $d_2$ values but distinct $p_\mathrm{epitope}$ values in the seasons 1996--1997 and 2004--2005 \citep{Gupta2006}.  Note that if $p_\mathrm{epitope}$ is incorporated into the linear models shown in Figure \ref{fig:d1} and \ref{fig:d2}, the $R^2$ value is increased. We fit a linear model $\mbox{vaccine effectiveness} = \alpha + \beta_1 p_\mathrm{epitope} + \beta_2 d_1 + \beta_3 d_2 + \epsilon$ in which $\epsilon$ is an error term. The fitted model is $\mbox{vaccine effectiveness} = 0.54 - 2.179 p_\mathrm{epitope} + 0.068 d_1 + 0.003 d_2$ with $R^2 = 0.72$.

\section{Discussion}
\label{sec:Discussion}

\subsection{Verification of the $p_\mathrm{epitope}$ Model}
\label{sec:pepitope_model}

Originally the $p_\mathrm{epitope}$ model was implemented for the
H3N2 virus, where $p_\mathrm{epitope}$ correlates with H3N2 vaccine
effectiveness with a significantly larger $R^2$ than do $p_{\text{all-epitope}}$ and $p_\mathrm{sequence}$ \citep{Gupta2006,Sun2006}.
In the case of H1N1, the advantage of $p_\mathrm{epitope}$ over
$p_{\text{all-epitope}}$ and $p_\mathrm{sequence}$ is not as remarkable as for H3N2. We speculate
that antibodies against the H3N2 virus may bind to a small fixed
region on the surface of H3 hemagglutinin while antibodies against
the H1N1 virus may have multiple binding regions available. In other
words, we speculate that the dominant epitope in H3 hemagglutinin may
contribute substantially to the escape of the H3N2 virus from host
antibodies, while escape mutations may occur in the dominant epitope as
well as perhaps the subdominant epitopes of H1 hemagglutinin.  Our speculation comes from the fact that the epitope region in H1N1 contains more amino acid positions than does that in H3N2 \citep{Deem2009}.

Two recent epidemiological studies \citep{Skowronski2010,MMWR2009b} present further support of the $p_\mathrm{epitope}$ model. Before the emergence of the H1N1 pandemic flu in April 2009, the 2008--2009 flu season was dominated by subtype H1N1 seasonal flu. Both the dominant circulating strain and the vaccine strain in the 2008--2009 season were A/Brisbane/57/2007 (H1N1) \citep{MMWR2009}. The observed vaccine effectiveness against seasonal flu was 44\% (95\% CI: 33\% to 59\%) \citep{Skowronski2010}. The $p_\mathrm{epitope}$ model predicts the vaccine effectiveness as 53\%, which falls into the 95\% CI of the reported vaccine effectiveness.

After April 2009, a new peak of influenza activity emerged. The dominant circulating strain in this period was the pandemic H1N1 strain A/California/7/2009 \citep{MMWR2009c,MMWR2009d}. The reported effectiveness of the 2008--2009 seasonal flu vaccine against the H1N1 pandemic flu was $-50\%$ to $-150\%$ \citep{Skowronski2010} and $-10\%$ (95\% CI: $-43\%$ to 15\%) \citep{MMWR2009b}. The value of $p_\mathrm{epitope}$ between A/California/7/2009 and A/Brisbane/57/2007 is 0.77 with epitope B as the dominant epitope. The vaccine effectiveness forecast by the $p_\mathrm{epitope}$ model is $-39\%$, which agrees with the measured vaccine effectiveness values.

\subsection{Comparison of H3N2 and H1N1 Vaccine Effectiveness and Evolution Rates}
\label{sec:Patterns}

The $p_\mathrm{epitope}$ model has been previously applied to the
prediction of H3N2 vaccine effectiveness \citep{Gupta2006}. The H3N2
vaccine effectiveness with $p_\mathrm{epitope}=0$ is 44.6\%, and vaccine
effectiveness is greater than zero for $p_\mathrm{epitope}<0.184$
\citep{Gupta2006}. Thus, H1N1 vaccines tend to have higher
vaccine effectiveness compared to H3N2 vaccines, as shown in Figure \ref{fig:pepitope}. The comparison between H3N2 and H1N1 vaccine effectiveness (Figure \ref{fig:pepitope} versus Figure 2 of \citep{Gupta2006}) illustrates that H1N1 vaccine has higher effectiveness than the H3N2 vaccine as a function of $p_\mathrm{epitope}$. This observation suggests that the host immune system is more effective at recognizing and eliminating the H1N1 virus ($p_\mathrm{epitope} = 0$), and that humoral cross immunity is stronger for H1 hemagglutinin ($p_\mathrm{epitope} > 0$). This observation also explains why an H3N2 epidemic is usually a more severe health threat than an H1N1 epidemic. We propose that H1N1 has a longer history of circulating in
the human population, so human immune system may recognize H1N1
more effectively, and this may be the reason that under stronger immune pressure, the H1N1 virus may have a higher degree of adaptation to the human host. In the following discussion, we verify this hypothesis by two facts. First, the H1N1 virus has a larger antigenic diversity than does the H3N2 virus. Second, the H1N1 virus presents higher evolutionary rate in the per dominant season basis.

\begin{figure}
\centering
\includegraphics[width=5in]{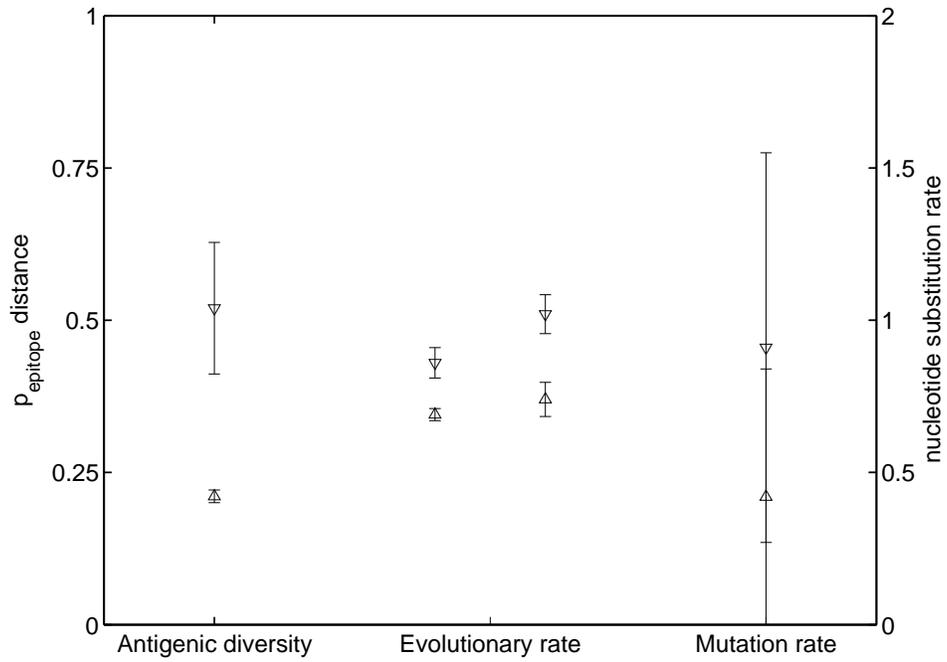}
\caption{The comparison between H3N2 (triangle up) and H1N1
(triangle down) in regard to the antigenic diversity, the evolutionary
rate between 1980 and 2000 (left), the evolutionary rate between
2000 to 2007 (right), and the mutation rate on a short time scale
without fixation. The antigenic diversity is measured with
$p_\mathrm{epitope}$, the unit of evolutionary rate is $10^{-3}$
nucleotide substitution/site/year, and the unit of mutation rate is
$10^{-6}$ nucleotide substitution/site/day.} \label{fig:compare}
\end{figure}

To compare the antigenic diversities of H1N1 and H3N2, we downloaded from the NCBI database on
13 August 2009 all the amino acid sequences of H3 hemagglutinin
collected in the 18 years with H3N2 dominant circulating strains
\citep{Gupta2006} and those of H1 hemagglutinin collected in 7 years
with H1N1 dominant circulating strains (Table \ref{tab:summary}).
Thus 18 subsets of H3N2 sequences and 7 subsets of H1N1 sequences
were formed. The centers of these subsets are the corresponding vaccine strains in the same season of the circulating virus.
The radius of each subset is obtained by the calculation of
$p_\mathrm{epitope}$. First, the strains with the top 5\%
$p_\mathrm{epitope}$ antigenic distance measure to the center of each subset were
selected, to focus on the extent of viral evolution. Second, the
$p_\mathrm{epitope}$ between these selected strains and the center
were averaged in each year as the radius. Third, the radii were
averaged over all the 18 years for H3N2 and over 7 years for H1N1.
That is, the average radius of the top 5\% was calculated in each
year. As a result, the average H3N2 subset radius with the vaccine
strains as the centers is 0.211. The average H1N1 radius is 0.520
with the vaccine strains as the centers. This difference between the
H3N2 radius and the H1N1 radius is significant with the $p$-value
0.0118 using the Wilcoxon rank-sum test. Consequently, the H1N1
virus has a larger antigenic diversity in each season
compared to the H3N2 virus, as shown in Figure \ref{fig:compare}.

We also compared the evolutionary rates of H1N1 and H3N2 because evolutionary rate of
the virus is an index of the selection pressure
of the virus. The virus undergoes less immune pressure in a non-dominant season and high immune pressure in a dominant season. It has been noticed that in H1 and H3 hemagglutinin, the region outside epitopes presents significantly lower evolutionary rate than do the epitopes \citep{Deem2009,Ferguson2003}. This phenomenon indicates that without immune pressure, the spontaneous evolutionary rates of both H1N1 and H3N2 are low. Therefore, a higher evolutionary rate of one virus subtype in a dominant season comes from the higher immune pressure rather than neutral evolution, and we reject the alternative scenario that the higher evolutionary rate causes a virus subtype to be dominant in one season. So the evolutionary rate per dominant season is a
natural measure of the virus evolution. Between 1983 to
1997, H3N2 was dominant in 8 of 15 years, and between 1977 to 2000,
H1N1 was dominant in 5 of 24 years \citep{Ferguson2003}.  Between 1980 to 2000, the
HA1 domain of H3 hemagglutinin has a higher annual evolutionary rate of
$3.7 \times 10^{-3}$ nucleotide substitution/site/year than does the HA1 domain of H1
hemagglutinin, which has the annual evolutionary rate of $1.8 \times 10^{-3}$ nucleotide
substitution/site/year \citep{Ferguson2003}.
Measured on a per dominant season basis, however, the HA1 domain of H1
hemagglutinin evolves faster in its dominant season with the rate of $8.6 \times
10^{-3}$ nucleotide substitution/site/dominant season than does the
H3 hemagglutinin with the rate of $6.9 \times 10^{-3}$ nucleotide
substitution/site/dominant season. The difference is significant
with a $p$-value 0.0008. Similarly, between 2000 and 2007, the HA1
domain of H1 hemagglutinin evolves faster in its dominant season with the rate of
$10.2 \times 10^{-3}$ nucleotide substitution/site/dominant season
than does the H3 hemagglutinin with the rate of $7.4 \times 10^{-3}$ nucleotide
substitution/site/dominant season. The difference is significant
with a $p$-value 0.0005 \citep{Zaraket2009}. Here we have divided
the annual evolutionary rate by the proportion of dominant years for
both H1 and H3 hemagglutinin. Even on a short time scale without
fixation, H1 hemagglutinin shows a comparable or higher mutation
rate of $9.1 \times 10^{-6}$ nucleotide substitution/site/day than H3
hemagglutinin of $4.2 \times 10^{-6}$ nucleotide
substitution/site/day ($p=0.26$) \citep{Nobusawa2006}, probably
caused by the adaptation to the higher immune pressure, at least for
some strains. To make this last point, we have assumed that the mutation rate of the HA
gene is the same as that of the NS gene. We assume that the same
polymerase is operating on these two genes, and so the mutation
rates are expected to be the same. The comparisons of evolutionary
rates and mutation rates between H3N2 and H1N1 are summarized in
Figure \ref{fig:compare}.

\subsection{The $p_\mathrm{epitope}$ Model as a Supplement to HI Assay}

For both H1N1 (this paper) and H3N2 \citep{Gupta2006}, the HI assay correlates less well with vaccine effectiveness than does $p_\mathrm{epitope}$. Collection of HI assay data measuring antigenic distance is also more time-consuming and more expensive compared to the $p_\mathrm{epitope}$ model.  Many hundreds of strains are circulating and collected in an average flu season, thus an HI table with tens of thousands of entries needs to be built to assess the antigenic distance between each pair of strains. With the high-throughput sequencing technology generating hemagglutinin sequence data, such antigenic distances are easily measured with the sequence-based antigenic distance measure $p_\mathrm{epitope}$, which correlates to a greater degree with vaccine effectiveness than do the HI data.

The $p_\mathrm{epitope}$ model is developed to provide researcher and health authorities with a new tool to quantify antigenic distance and design the vaccine. We do not suggest that $p_\mathrm{epitope}$ should substitute for the current HI assay, but rather suggest that $p_\mathrm{epitope}$ serves as an additional assessment when selecting vaccine strains. Using $p_\mathrm{epitope}$ to supplement to HI assay data may allow
researchers and health authorities to more precisely quantify the
antigenic distance between dominant circulating strains and candidate vaccine
strains. The adoption of the $p_\mathrm{epitope}$ theory may also
allow researchers to minimize the cost and the number of ferret
experiments and to correct HI assay data in some situations.

\paragraph{Acknowledgements} Keyao Pan's research was supported by
the Gulf Coast Consortia Nanobiology Training Program (NBTP).
Krystina C. Subieta's work at Rice University was funded by the HHMI
Summer Undergraduate Research Internship Program in
Bionanotechnology. This project was also partially supported by
DARPA grant HR 00110510057.

Edited by Gideon Schreiber

\bibliographystyle{peds}
\bibliography{references}

\end{document}


\title{Supplementary Materials}
\author{$\mbox{Keyao Pan}^1$, $\mbox{Krystina C. Subieta}^3$, and
$\mbox{Michael W. Deem}^{1,2}$\\\\
Department of $^1$Bioengineering and $^2$Physics \& Astronomy, Rice University,\\
6100 Main Street, Houston, TX 77005\\
$^3$Department of Mechanical and Aerospace Engineering, University of Florida,\\
Gainesville, FL 32611}

\date{}

\maketitle

\section{Humoral Immune System Plays a Major Role in Immunity to Influenza}
\label{sec:Humural}

The influenza vaccine considered in this study is the trivalent inactivated vaccine (TIV) administered by intramuscular injection. The effective components of TIV are hemagglutinin (HA) and neuraminidase (NA) that noticeably induce the humoral immunity but activate the cellular immunity less vigorously \citep{Doherty2008}. The other, cold--adaptive trivalent live attenuated influenza vaccine (LAIV) is also believed to induce the cellular immunity to a low level \citep{Doherty2008}.

The humoral immunity greatly relies on the antigenic distance between the hemagglutinin of the vaccine and that of the dominant circulating strain. On the other hand, the cellular immune system focuses on the highly conserved internal proteins, which are the Matrix protein 1 (M1) and the nucleoprotein (NP) \citep{Lee2008}. In contrast to the antibodies, CD8+ and CD4+ T cells show notable cross immunity to a wide variety of strains \citep{Lee2008}. Like the cellular immune system, the antigen-unspecific innate immune system generates a homogeneous immune reaction against different influenza strains \citep{Janeway2005}.

For all these reasons ferret antisera, in which antibodies are the major immune component, is used in the hemagglutination inhibition (HI) assay as the conventional way to measure the antigenic distance between the vaccine strain and the dominant circulating strain. Therefore, we consider the antibody rather than the cellular or innate immune system to be the dominant element in our quantification of antigenic distance between two influenza strains and the key factor for influenza vaccine effectiveness.

\section{Evaluation of Vaccine Effectiveness}
\label{sec:Evaluation_of_Vaccine_Effectiveness}

By definition, vaccine efficacy is measured by controlled trials with initially susceptible subjects, while vaccine effectiveness is measured by epidemiological observance of susceptible population without giving placebo \citep{Kelly2009}. Vaccine efficacy is relatively more idealized than vaccine effectiveness, because vaccine effectiveness depends on vaccine efficacy and other environmental factors \citep{Torvaldsen2002}. Although the terms vaccine efficacy and vaccine effectiveness are interchangeable to some extent \citep{Torvaldsen2002}, we use the term vaccine effectiveness because factors other than vaccine strain and dominant circulating strain are involved in the studies used in our metaanalysis. The data source for vaccine effectiveness calculation used ILI as the primary endpoint, and studies we use contained controlled unvaccinated groups.

The data from four studies require additional clarification. The paper by Edwards et al.\ \citep{Edwards1994} did not provide the retrospectively reported influenza--like illness data prior to the 1987--88 season, so we use the number of ill subjects presenting for throat culture to calculate the morbidity rate $u$ and $v$. Note that in this study, subjects with influenza--like disease were required to show up for a throat culture, and when characterized by vaccination status the numbers of such patients is thus a reasonable estimation of the illness rate $u$ and $v$. Moreover, in other seasons when both retrospective data and number of presenting ill subjects were available, the vaccine effectiveness calculated from retrospective data and numbers of presenting ill subjects are similar to each other, especially for subtype H1N1 \citep{Edwards1994}.
In Grotto et al.'s study \citep{Grotto1998} using influenza strains from Israel in the 1995--96 season, the numbers of sampled H1N1 and H3N2 strains in Israel was 7 and 35, respectively. The samples were collected from six clinics in December that was in the middle of the influenza season. However, the number of both subjects and viruses sampled are limited. At the global level with more available data, it was observed that H1N1 and H3N2 were co-circulating with comparable frequencies, and H1N1 virus was found in North America and part of Eurasia including Israel \citep{MMWR1996}. The proportion of H1N1 in samples during the 1995--96 season in USA ranks \#5 in H1N1 proportion since 1977 \citep{Ferguson2003}. In the same season, H3N2 vaccine strain and dominant circulating strain were a perfect match, so the decrease in the overall vaccine effectiveness is expected to be due to the mismatch in the H1N1 component. Therefore we treat H1N1 here as a co-circulating strain and take into account the vaccine effectiveness reported in this article \citep{Grotto1998}. Keitel et al. \citep{Keitel1997} reported that the dominant circulating strain in the 1983--84 season was A/Chile/1/83 rather than A/Victoria/7/83 in this table and in other cited studies. The illness rate $u$ and $v$ are small, and so the standard error of vaccine effectiveness is 64.8\%, which is unacceptable. Thus the use of Keitel et al.'s data for 1983--84 season is not appropriate to the vaccine effectiveness assessment. The reference by Couch et al.\ \citep{Couch1996} did not provide original data $n_u$, $N_u$, $n_v$, and $N_v$ for the calculation of vaccine effectiveness. Error bars of vaccine effectiveness in these seasons were calculated with other data sources.

\section{Robustness of the $p_\mathrm{epitope}$ Model}
\label{sec:Rationale}

\begin{figure}
\centering
\includegraphics[width=5in]{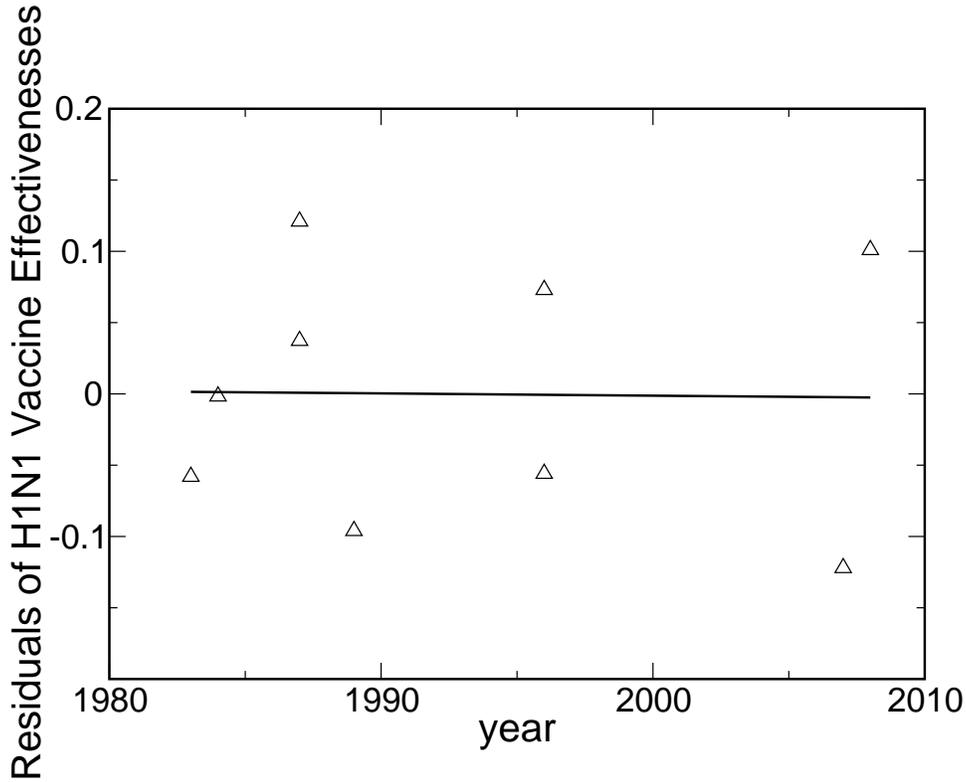}
\caption{The linear regression with $R^2=0.0003$ of the residuals of
H1N1 vaccine effectiveness versus year. Data from Table 1 and Figure 2 in the main text. The slope of the trend line is
$-0.0002$/year. ANOVA test: $H_0$: $\mbox{slope}=0$, $F=0.0021$, and
$p=0.96$. The null hypothesis that these residuals are independent
of time cannot be rejected.} \label{fig:residual_H1}
\end{figure}

\begin{figure}
\centering
\includegraphics[width=5in]{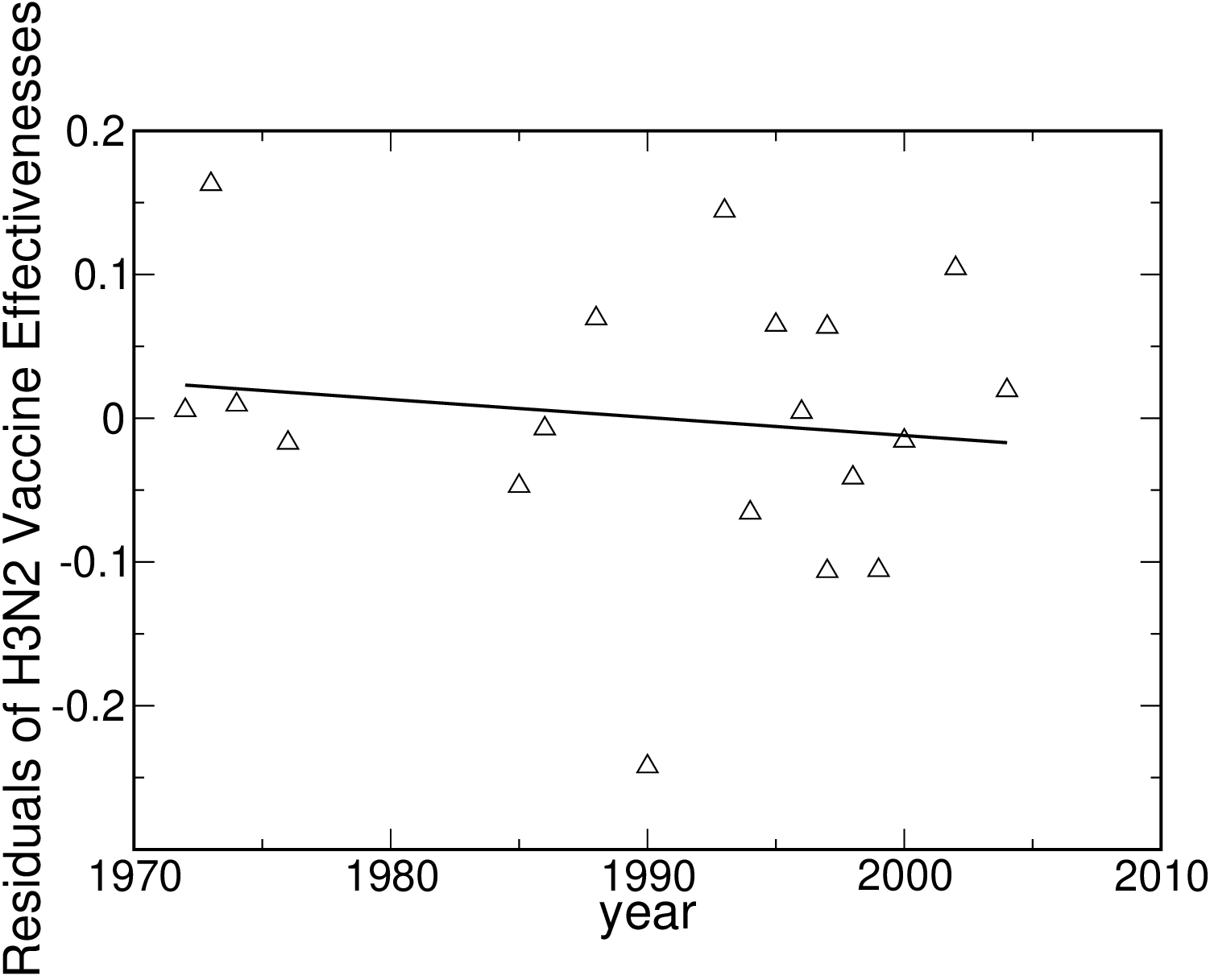}
\caption{The linear regression with $R^2=0.018$ of the residuals of
H3N2 vaccine effectiveness versus year. Data from \citep{Gupta2006}. The slope of the trend line is
$-0.0013$/year. ANOVA test: $H_0$: $\mbox{slope}=0$, $F=0.32$, and
$p=0.58$. The null hypothesis that these residuals are independent
of time cannot be rejected.} \label{fig:residual_H3}
\end{figure}

Influenza vaccine effectiveness may depend not only on the
antigenic distance between the vaccine strain and the dominant circulating
strain quantified by $p_\mathrm{epitope}$, but also on the
percentage of people vaccinated, the time of vaccination in the
influenza season, influenza virus transmissibility and reproduction
rate, and individual's immune history. Thus, development of the
public health system and a greater fraction of the population being
vaccinated may result in a trend of both H1N1 and H3N2 vaccine
effectiveness. The statistics of vaccine effectiveness could be biased by these factors. Nevertheless, greater than 50\% of the H1N1 and H3N2 vaccine effectiveness
are explained by $p_\mathrm{epitope}$, since $R^2 > 1/2$. To show
that the model of vaccine effectiveness can be well reduced to a linear
form between $p_\mathrm{epitope}$ and vaccine effectiveness, we
calculated the residuals of linear regression of vaccine effectiveness on
$p_\mathrm{epitope}$, and performed
another linear regression of these residuals versus year. The trend
line of the residuals has a slope of $-0.0002$/year and the null
hypothesis that the slope equals zero is not rejected $(p=0.96)$, as
shown in Figure \ref{fig:residual_H1}. The residuals of H3N2 vaccine
effectiveness \citep{Gupta2006} were also correlated with the year and
the slope $-0.0013$/year is not significantly different with zero
$(p=0.58)$, as shown in Figure \ref{fig:residual_H3}. Therefore the
contribution of other simple time--dependent factors other than
$p_\mathrm{epitope}$ to H1N1 and H3N2 vaccine effectiveness in humans is
negligible. Our analysis suggests that the vaccine effectiveness data in this paper are negligibly affected by these potential biases.

Despite the limited number of available data points in this study, the correlation line between the $p_\mathrm{epitope}$ and the vaccine effectiveness has statistical meaning. In Figure 2 in the main text, the trend line is $\mbox{vaccine effectiveness} = -1.19 \ p_\mathrm{epitope} + 0.53$, the greatest determinant of which is the data point 1986--87 (b). If this data point is removed, the trend line becomes $\mbox{vaccine effectiveness} = -1.63 \ p_\mathrm{epitope} + 0.54$, which is not fundamentally distinct with the original trend line. In the data point 1986--87 (b), the difference of the vaccine effectiveness predicted by these two models is 0.13, which is roughly one standard error. In reality, most $p_\mathrm{epitope}$ values are less than 0.1, and so most of the differences between these two predicted vaccine effectiveness values are less than 0.034, which is within the noise levels of the epidemiological measurements.

The basis for calculating $p_\mathrm{epitope}$ is a set of well
defined epitopes. An early definition of the five epitopes in H1
hemagglutinin \citep{Caton1982} did not identify numerous amino acid
positions in which mutations were frequently selected in history. These
positions are presumably under strong antigenic pressure to be selected for
escape mutation. The more recent definition of H1 epitopes
incorporates these additional amino acid positions as well as amino
acids from the epitopes of H3 hemagglutinin \citep{Deem2009}. Likely
additional experiments on the H1 epitopes will allow
further refinement of the calculation of $p_\mathrm{epitope}$. Only nine
epidemiological data points are available since the reemergence of
H1N1 virus in humans in 1977. The $p_\mathrm{epitope}$ model
parameters may be further improved as epidemiological data are
accumulated.

The antigenic properties are determined by a small number of amino acid substitutions, because the positions and the amino acids introduced by mutation have distinct effects on the change of antigenic distance between vaccine strain and dominant circulating strain. For example, mutations yielding charged amino acids in the dominant epitope are favorable for the virus, and may be the key amino acid substitution for the antigenic properties (K. Pan et al., submitted). An improved sequence--based model might assign different amino acid substitutions with weights determined by the decrease of binding constant between HA and antibody using free energy calculation (K. Pan and M. W. Deem, submitted). With the current knowledge, the less precise but safe $p_\mathrm{epitope}$ model assigns the amino acid substitutions in the dominant epitope with weight one, and assigns other amino acid substitutions with weight zero. The $p_\mathrm{epitope}$ model can nevertheless correlate with the vaccine effectiveness better than the antisera data.   That is, for both H1N1 and H3N2, the $p_\mathrm{epitope}$ method is superior to other methods in current use.

\section{Comparison of the $p_\mathrm{epitope}$ Model and the HI Assay for H3N2 Virus}
\label{sec:Comparison_pepitope_HI}

In some cases $p_\mathrm{epitope}$ model detects antigenic variants better than the HI assay. In the 2003--04 Northern hemisphere flu season, the majority of isolated H3N2 strains were similar to A/Fujian/411/2002 using HI assay \citep{WHOWER2004}, hence WHO recommended a A/Fujian/411/2002-like strain as the 2004--05 Northern hemisphere H3N2 vaccine component, and A/Wyoming/3/2003 was selected. Although A/Wyoming/3/2003 is similar to A/Fujian/411/2002 circulating in 2004--05 ("antigenically equivalent" by HI data \citep{Harper2004}), the vaccine effectiveness was only moderate \citep{Gupta2006}. Interestingly, the $p_\mathrm{epitope}$ between A/Fujian/411/2002 and A/Wyoming/3/2003 was also moderate ($p_\mathrm{epitope} = 0.095$), predicting the moderate vaccine effectiveness. In fact, the $p_\mathrm{epitope}$ method can also detect antigenic variants more rapidly as they emerge \citep{He_submitted}.

\bibliographystyle{peds}
\bibliography{references}